# Relevance Score of Triplets Using Knowledge Graph Embedding

## The Pigweed Triple Scorer at WSDM Cup 2017

Vibhor Kanojia*
Yahoo! Japan
vkanojia@yahoo-corp.jp

Riku Togashi*
Yahoo! Japan
rtogashi@yahoo-corp.jp

Hideyuki Maeda*
Yahoo! Japan
hidmaeda@yahoo-corp.jp

## ABSTRACT

Collaborative Knowledge Bases such as Freebase [1] and Wikidata [2] mention multiple professions and nationalities for a particular entity. The goal of the WSDM Cup 2017 [3] Triplet Scoring Challenge was to calculate relevance scores between an entity and its professions/nationalities. Such scores are a fundamental ingredient when ranking results in entity search. This paper proposes a novel approach to ensemble an advanced Knowledge Graph Embedding Model with a simple bag-of-words model. The former deals with hidden pragmatics and deep semantics whereas the latter handles text-based retrieval and low-level semantics.

## 1. TASK INTRODUCTION

Many entities usually have multiple professions or nationalities, and it is often desirable to rank the relevance of these individual triplets. The goal of the challenge was to compute a score in the range [0, 7] that measures the relevance of the statement expressed by the individual triplet compared to other triplets from the same relation. Participants were provided with a list of 385,426 entities along with five files,

- *profession.kb*: all professions for a set of 343,329 entities
- *nationality.kb*: all nationalities for a set of 301,590 entities
- *profession.train*: relevance scores for 515 tuples (pertaining to 134 entities) from profession.kb
- *nationality.kb*: relevance scores for 162 tuples (pertaining to 77 entities) from nationality.kb
- *nationality.kb*: 33,159,353 sentences from Wikipedia with annotations of the 385,426 entities

Apart from these, the participants were allowed to use any kind or amount of additional data (except for human/judgements). The output of this task was to generate relevance scores for all the triplets, 0 being the lowest relevance, and 7 being the highest.

*Equal Contribution

## 2. OUR APPROACH

We used a pipeline based ensemble model, the system diagram of which is shown in Figure 1.

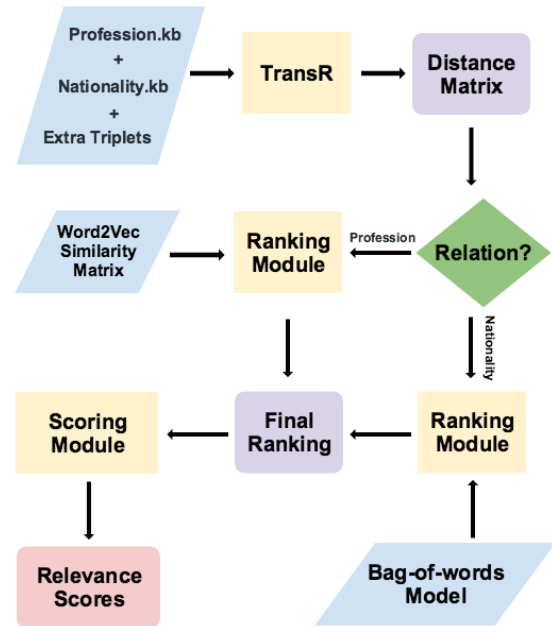

**Figure 1: Flow chart of our Approach**

## 2.1 Step 1: TransR

The first part of our approach makes use of a shallow learning technique for Knowledge Graph Embedding, TransR [4]. TransR is a translation-based embedding model which, given a fact of a knowledge base, represented by a triplet $\triangle(h, r, t)$ where $h$, $r$, $t$ indicate a head entity, a relation and a tail entity respectively. aims to learn embedding vectors $h$, $r$ and $t$.

In TransR, for each triplet $\triangle(h, r, t)$, entity embeddings are set as $h, t \in \mathbb{R}^k$ and relation embedding is set as $r \in \mathbb{R}^d$. Note that the dimensions of entity embeddings and relation embeddings are not necessarily identical, i.e., $k \neq d$. For each relation $r$, TransR sets a projection matrix $M_r \in \mathbb{R}^{k \times d}$, which may project entities from entity space to relation space. With the mapping matrix, TransR defines the projected vectors of entities as,

$$h_r = hM_r, t_r = tM_r \quad (1)$$

TransR defines the score function as,

$$f_r(h, t) = \|h_r + r - t_r\|_2^2 \quad (2)$$

Training phase uses margin-based ranking loss to encourage discrimination between golden triplets and incorrect triplets:

$$\mathcal{L} = \sum_{(h,r,t)\in \triangle} \sum_{(h',r',t')\in \triangle'} [f_r(h,t) + \gamma - f'_r(h',t')]_+ \quad (3)$$

where $[x]_+ \triangleq max(0,x)$, $\triangle$ is the set of positive(golden) triplets, $\triangle'$ denotes the set of negative triplets generated by randomly shuffling heads or tails in the triplets $\triangle(h,r,t)$. This step is known as *Negative Sampling*. $\gamma$ is the margin separating positive and negative triplets.

The role of TransR in our approach is to learn embeddings of entities such that entities related to each other will be closer in the euclidean space. The approach is similar to using a *Word2Vec* [5] model except that TransR uses triplets data as training input.

In order to achieve semantically sound embedding, we introduced some more triplets data using the Wikidata knowledge base. Apart from the triplets dataset provided, we extracted triplets for relations such as *place_of_birth*, *place_of_death*, *employer*, *country*, *capital*, etc. Such relations are closely related to the relations *profession* and *nationality*. The idea was to provide the model with more knowledge about the entities so that it can learn a semantically well organised vector embedding.

We used the trained embeddings to rank all the triplets in the training dataset using the score function $f_r(h,t)$. Profession/Nationality more closely related to a particular entity had relatively smaller value of the score function as compared to other professions/nationalities.

## 2.2 Step 2: Ranking Module

Keeping in mind that the scores were crowdsourced using human judgement and analysing the training files, we made the following observations,

- Similar profession pairs, such as *[Actor, Model]* or *[Singer, Composer, Songwriter]*, had similar scores in the sample data provided

- In case of multiple nationalities, scores were biased towards the birth country

- TransR predicted the most relevant profession/nationality of a particular entity correctly, but had difficulty ranking the other, not so relevant, professions/nationalities

### 2.2.1 Profession Ranking: Word2Vec

In order to modify the profession ranking based on the observations, we used an already trained *Word2Vec* model to generate 200x200 similarity distance matrix for all the given professions. We used a naive threshold based linear combination to modify the ranking obtained by TransR.

Based on observation 3, we improved the ranking of professions which were similar to the most relevant profession of an entity using the *Word2Vec* similarity matrix. For example, according to TransR, *A. R. Rahman's* most relevant profession was *singer-songwriter*. So we improved the ranking of professions similar to *singer-songwriter*, such as *singer* or *composer*.

Let us consider a head entity $h$ having $n$ professions [$p_1$, $p_n$] where $p_i$ has rank $i$ after Step 1. We used the following equation to update the ranking,

$$NewRank(p_i) = max(i - d, 1) \quad (4)$$

where,

$$d = \begin{cases} 2 & 0 < SD(i,1) \leq 0.2 \\ 1 & 0.2 < SD(i,1) \leq 0.4 \\ 0 & 0.4 < SD(i,1) \leq 0.7 \\ -1 & 0.7 < SD(i,1) \leq 1 \end{cases}$$

,
SD(i, j) = Similarity Distance between $p_i$ and $p_j$.

For example, for the entity *Bob Dylan*, the most relevant profession was *Singer-songwriter*. In Step 2, *Word2Vec* enabled our model to increase the ranking of the professions *Lyricist*, and *Guitarist* and decrease the ranking of the professions *Author*, and *Actor*. Note that we may get multiple tails with same rank after Step 2 and hence, the same score, which can be observed in the training data as well.

### 2.2.2 Nationality Ranking: Bag-of-words

In order to modify the nationality ranking, we used a simple bag-of-words model created from the wiki-sentences list provided to the participants.

We observed that demonyms are a useful way to identify the nationality of an entity. We created a list of 100 country-demonym pairs, and created a bag-of-words model for each entity and those 100 nationalities, replacing demonyms by country names on the way. For each entity, we picked the nationality with the highest count, given that count is at least 3, and assigned rank 1 to that nationality.

Instead of using a linear combination, we gave priority to Bag-of-words model for nationality ranking.

## 2.3 Step 3: Scoring Module

The competition defined accuracy as the percentage of triplets for which the score computed by our system differs from the ground truth by at most 2. Keeping this in mind we decided to assign scores to triplets in the range [2, 5]. The scoring module was a simple step down model,

$$\begin{aligned} Rank &= 1 \implies Score = 5 \\ Rank &= 2 \implies Score = 3 \\ Rank &\geq 3 \implies Score = 2 \end{aligned} \quad (5)$$

## 3. OTHER APPROACHES

This section mentions some of the other approaches that we tried for the task of Triplet Scoring but didn't use in the final software:

- **Word2Vec:** Entities from neigbouring countries may have similar name-styles or cultures. A few such pairs are *[India, Pakistan]*, or *[Japan, China, Taiwan]*. We observed that such countries had similar scores in the training data for nationality ranking. We decided to use another *Word2Vec* model which could find similarity between such country pairs and assign scores accordingly. This approach required a new *Word2Vec* model to be trained on a carefully filtered dataset.

- **Bag-of-words:** We tried using a bag-of-words model for profession ranking similar to nationality ranking but a simple bag-of-words model was not able to fully understand the complicated sentence structure, and as result, the triplets obtained were of low quality. An indicator word sometimes co-occurs with the entity in question, but is actually related to another person. For example, *Barack Obama* is listed as *Author* which is certainly not his primary profession, but the word *Author* is mentioned frequently with *Barack Obama* which, in reality, is referring to another person writing a book about *Obama*.

- **Twitter Data:** Twitter data is an excellent source for text mining and knowledge completion tasks. We tried using a bag-of-words model created from Twitter data but the text was mostly garbled and full of noise due to character limit. In order to reduce the noise and filter only relevant information, we clustered most of the entities into 12 clusters based on the month in which they were born, and trained separate bag-of-words models, but the results were unsatisfactory. For example, tweets in March had various tweets about *Albert Einstein*, but due to character limit there were very few occurrences of *Theoretical Physicist*, and many people had tweeted *Einstein's* quotes/stories containing words like *Teacher* and *Mathematician*, which further deteriorated the bag-of-words model.

- **Wikidump:** In this approach, we trained a softmax classifier that classifies the bag-of-words data, calculated from Wikidump, for each entity into 100 professions. The output of this classifier was a vector of posterior probability. These probability vectors were used as teacher signals (soft targets). Due to generalisation capability of the classifier, the probabilities of professions frequently appearing together with the entity in the Wikidump were relatively high. In other words, the output of softmax layer can be seen as probability distribution of that entity's profession into 100 professions. Due to variations in the document size for each entity, the results of this approach also varied.

## 4. EVALUATION RESULTS

The triplet scoring task was evaluated on three measures,

- **Average score difference(ASD):** for each triplet, take the absolute difference of the relevance score computed by the system and the score from the ground truth; add up these differences and divide by the number of triplets.

- **Accuracy:** the percentage of triplets for which the score computed by the system differs from the ground truth by at most 2.

- **Kendall's Tau:** for each relation, for each subject, compute the ranking of all triplets with that subject and relation according to the scores computed by the system and the score from the ground truth. Compute the difference of the two rankings using Kendall's Tau.

The task was evaluated on a test set of 710 triplets, *profession*: 513 triplets and *nationality*: 197 triplets. Table 1 shows the performance of our model over these various evaluation metrics. Our team, *Pigweed*, secured 11th position out of a total of 21 teams. As seen in the table, due to restriction of scores in the range [2, 5] to boost the accuracy, our model suffered in the other evaluation metrics.

**Table 1: Evaluation Results**

|  | Accuracy | ASD | Tau |
|---|---|---|---|
| Bokchoy(1st Position) | 0.87 | 1.63 | 0.33 |
| Pigweed(Score: [2, 5]) | 0.74 | 1.94 | 0.48 |
| Pigweed(Score: [0, 7]) | 0.73 | 1.91 | 0.43 |

For experiments on TransR, we used Adam Optimiser with initial learning rate $\lambda = 0.001$, batch size = 256, dimension of vectors d = 50, dimension of matrix k = 50, varied the margin $\gamma$ in the range = {0.2, 0.5, 1.0, 2.0}, and used $L_2$ distance between vectors. We trained the model on a dataset of 983,683 triplets for 2000 epochs.

We enforced constraints on the norms of the embeddings $h$, $r$, $t$ and the mapping matrices, i.e. $\forall\ h, r, t$, we have $\|h\|_2 \leq 1$, $\|r\|_2 \leq 1$, $\|t\|_2 \leq 1$, $\|hM_r\|_2 \leq 1$, $\|tM_r\|_2 \leq 1$.

We used an already trained *Word2Vec* model which contains 300-dimensional vectors for 3 million words and phrases. We assigned similarity distance = 0.5 for nationalities and professions which were not present in the word-list.

## 5. CONCLUSION

In this paper, we discuss the approaches that we ventured for the Triple Scoring Challenge [6] at WSDM Cup 2017. Our idea was to challenge this task using Knowledge Graph Embedding techniques. Our final model secured 11th position(out of 21) achieving an accuracy of 74% on a test dataset of 710 triplets.

During the task, we realised that simple techniques such as bag-of-words performed much better than *Word2Vec* when we consider *nationality*. However, for *profession* relation *Word2Vec* outperformed simple models. We reckon that this is because, given a sentence, it is quite simple to figure out the nationality of a person using demonyms knowledge whereas finding the profession is a challenging task which requires high level NLP techniques.

We also wanted to discover some new use cases of TransR, and with some modifications in the original algorithm like adding relevance score information in the loss equation, its performance can be improved. We used very naive scoring module due to lack of time and in order to keep things simple, which could be further improved for better results. Using a simple scoring module improved the accuracy but resulted in a high Tau value.

We used an already trained *Word2Vec* model which was very effective in finding similar professions. However, a *Word2Vec* model trained on filtered data centred around the given entities(from Twitter or Wikipedia) can further improve the performance of the model. Refer to *this paper* [7] which discusses the approach used by various teams and their in-depth analysis.